# Buffered Aloha with *K*-Exponential Backoff
# Part I: Stability and Throughput Analysis

Tony T. Lee, *Fellow, IEEE*, and Lin Dai, *Member, IEEE*

*Abstract*—This two-part paper series studies the performance of buffered Aloha networks with *K*-Exponential Backoff collision resolution algorithms. Part I focuses on stability and throughput analysis and Part II presents the delay analysis.

In Part I, the buffered Aloha network is modeled as a multi-queue single-server system. We adopt a widely used approach in packet switching systems to decompose the multi-queue system into independent first-in-first-out (FIFO) queues, which are hinged together by the probability of success of head-of-line (HOL) packets. A unified method is devised to tackle the stability and throughput problems of *K*-Exponential Backoff with any cutoff phase *K*. We demonstrate that a network with *K*-Exponential Backoff can be stabilized if the retransmission factor *q* is properly selected. The stable region of *q* is characterized and illustrated via examples of Geometric Retransmission (*K*=1) and Exponential Backoff (*K*=∞). With an increasing number of nodes *n*, we show that the stable region of Geometric Retransmission rapidly shrinks, and vanishes as *n*→∞. In contrast, the stable region of Exponential Backoff does not vary with the network population *n*, implying that a stable throughput can be achieved in networks with Exponential Backoff even with an infinite number of nodes. All the analytical results presented in this paper series are verified by simulations.

## I. INTRODUCTION

A fundamental problem of multi-access communications is how to efficiently share the channel resource among multiple users. From the Aloha network to today's IEEE 802.11 Wi-Fi network, random access has proven to be a simple yet elegant solution; transmit if there is a request, and back off if a collision occurs. The minimum coordination and distributed control have enabled random access to become one of the most widely deployed network technologies used today [1].

Despite the huge success in practical systems, the performance of even the simplest version of random access, Aloha [2-4], is not clearly understood yet. Throughput analysis of random access networks can be traced back to Abramson's landmark paper [2]. Assuming an infinite number of nodes, Abramson proposed to model the aggregate traffic as a Poisson random variable with parameter *G*, which captures the essence of contentions among users without intricate analysis. The steady-state equilibrium throughput $G\exp(-G)$ derived from this simplified model sheds useful insight into many aspects of network performance, such as the maximum network throughput of $e^{-1}$ when $G=1$.

Research interests have bifurcated since then. On one hand, intense activities have focused on the stability analysis of Abramson's Aloha protocol under the assumption that the network is saturated and each node always has a packet to transmit. Most of these studies ignored queueing aspects, and concentrated on the throughput of the entire network.

Specifically, early work showed that slotted Aloha is unstable with an infinite population [5-6]. To stabilize a finite-node Aloha system, a drift approach was developed to analyze the system transition so that the retransmission probability could be adjusted accordingly. This requires the knowledge of the number of backlogged nodes [7-10]. Rather than investigating how to estimate the backlogs [11-12], Binary Exponential Backoff (BEB) algorithms were proposed to achieve stability by reducing the retransmission probability according to the number of collisions the packet has experienced [13]. It was proven in [14] that BEB is also unstable with an infinite number of nodes. Later, under a finite-node model, it was shown that BEB can be stable if the aggregate arrival rate is sufficiently small [15]. Different upper bounds of the aggregate arrival rate have been developed since then [16-17] but none of them has become the consensus [18]. The stability issue of random access protocols remains an open problem.

In the meantime, a great deal of effort has been made to establish a buffered Aloha model, which was initiated in [19] and further developed in [20-25]. With the interactions among the different queues taken into consideration, an *n*-node buffered Aloha system was modeled as an *n*-dimensional random walk and the exact rate region for the two-node case was derived in [19]. Unfortunately, the generalization of this approach to an arbitrary *n*-node system encountered tremendous difficulties [20-24]. Note that progress has been made recently in characterizing the *n*-node rate region by assuming independence among nodes [25]. It was shown that the rate region is asymptotically exact when the number of users grows large. Nevertheless, the problem still seems intractable when Exponential Backoff is further involved.

The above inconsistency and confusion on the stability analysis of Aloha networks mainly originates from the various definitions of stability based on different analytical models adopted in previous studies. In fact, for buffered Aloha networks, it is a nature way to define stability from the system aspect. In this paper series, we define that a homogeneous *n*-node buffered Aloha network is stable if:
*1) the network throughput equals the aggregate input rate, or,*
*2) the mean queueing delay of input packets is finite.*
The above two definitions are referred to as throughput stability and delay stability, respectively. It is obvious that the delay stability implies the throughput stability, but the reverse is not necessarily true. A network that is stable in terms of throughput but fails to achieve delay stability is called *quasi-stable*.

In this paper, the buffered Aloha network is modeled as a multi-queue single-server system, where each node is equipped with an infinite buffer and treated separately. A widely used approach in packet switching systems [26-28] is adopted to decompose the multi-queue system into independent FIFO queues with Bernoulli arrivals of rate $\lambda$ packets per time slot. These decomposed Geo/G/1 queues are then hinged together by the service time that is determined by the probability of success *p* of Head-of-Line (HOL) packets.

It will be shown that the probability of success of a buffered Aloha network with *K*-Exponential Backoff has one *desired stable point* at $p_L$, one *unstable equilibrium point* at $p_S$ and one *undesired stable point* at $p_A$. Both throughput and delay stabilities are achievable at the desired stable point $p_L$ which is solely determined by the aggregate input rate $\hat{\lambda}$. At the undesired stable point $p_A$, however, the network throughput may depend on the backoff parameters such as the retransmission factor *q* and the cutoff phase *K*. For Geometric Retransmission (*K*=1), the network is always unstable when operating at the undesired stable point $p_A$. In contrast, throughput stability can still be achieved by Exponential Backoff (*K*=∞) networks, yet the delay performance is severely penalized and the network will become quasi-stable.

Our analysis also reveals that the retransmission factor *q* should be carefully selected to achieve stability. We characterize the *absolute-stable region* of *q*, inside which the network is guaranteed to operate at the desired stable point $p_L$. Outside the absolute-stable region, the network has a risk of evolving into the undesired stable point $p_A$. The *quasi-stable region* of Exponential Backoff is derived when the network is operating at $p_A$. Table I summarizes the major results for Geometric Retransmission and Exponential Backoff. Note that if throughput stability is the only concern, the complete stable region is the union of absolute-stable region and quasi-stable region.

A number of results on the stability issue reported in previous studies can be confirmed by our analysis. For example, we can see from Table I that the stable region of Geometric Retransmission is empty when the number of nodes *n* is infinite. This result agrees with [5-6] that the network with Geometric Retransmission is unstable as *n*→∞. For Exponential Backoff, it was proved in [29] that Binary Exponential Backoff (*q*=1/2) is unstable when the aggregate input rate $\hat{\lambda}$ is larger than ln2, but the issue remains open for the case of $\hat{\lambda} \leq$ln2. Our analysis shows that it can be stabilized if $\hat{\lambda} \leq \frac{1}{2}\ln 2$.

Beyond that, our analysis further presents the entire stable operating range of the network. The characterization of the stable region of retransmission factor *q* provides guidelines for protocol design and network management in practice. For given traffic input rate $\hat{\lambda}$ and network population *n*, appropriate values of *q* can be selected to achieve a stable throughput. Moreover, the offered load of each input buffer, $\rho$, is explicitly represented as a function of backoff parameters such as the retransmission factor *q* and the cutoff phase *K*. In contrast to the previous saturation analysis where $\rho$ is always assumed to be 1, the buffered model enables us to further explore the delay performance. The detailed delay analysis will be presented in Part II of the paper series.

The remainder of this paper is organized as follows. Section II establishes the network model and presents the preliminary analysis of buffered Aloha with backoff scheduling. The absolute-stable and quasi-stable regions are characterized Sections III and IV, respectively. Simulation results are provided in Section V and conclusions are summarized in Section VI.

Table I. Stable Region and Maximum Stable Throughput of Geometric Retransmission (*K*=1) and Exponential Backoff (*K*=∞)

|  | Geometric Retransmission (*K*=1) | | Exponential Backoff (*K*=∞) | |
| --- | --- | --- | --- | --- |
|  | Stable Region of Retransmission Factor *q* | Maximum Stable Throughput | Stable Region of Retransmission Factor *q* | Maximum Stable Throughput |
| Absolute-Stable: Network operates at the desired stable point $p_L$. | $S_L^{Geo} = \left[ \frac{\hat{\lambda}(1-p_L)}{p_L(n-\hat{\lambda})}, -\frac{\ln p_S}{n} \right]$ | $e^{-1}$ (with *q*=1/*n*) | $S_L^{Exp} = \left[ \frac{1-p_L}{1-\hat{\lambda}/n}, -\frac{\ln p_S}{n} \right]$ | ln*n*/*n* (with *q*=ln*n*/*n*) |
| Quasi-Stable: Network operates at the undesired stable point $p_A$ and throughput stability is achieved. | $S_A^{Geo} = \varnothing$ | Not defined | $S_A^{Exp} = [1-p_L, 1-p_S]$ | $e^{-1}$ (with *q*=1-$e^{-1}$) |
| Throughput stability is achieved. | $S^{Geo} = \left[ \frac{\hat{\lambda}(1-p_L)}{p_L(n-\hat{\lambda})}, -\frac{\ln p_S}{n} \right]$ | $e^{-1}$ (with *q*=1/*n*) | $S^{Exp} = [1-p_L, 1-p_S]$ for large *n* | $e^{-1}$ (with *q*=1-$e^{-1}$) |

The main notations used in this paper are listed as follows:
*n*: number of nodes
*K*: cutoff phase
*q*: retransmission factor (0<*q*<1)
*λ*: input rate per node
$\hat{\lambda}$: aggregate input rate. $\hat{\lambda}=n\lambda$
*ρ*: offered load of each node's queue
*p*: probability of success of each HOL packet
*G*: attempt rate
$p_L$: desired stable point. $p_L = \exp\{W_0(-\hat{\lambda})\}$
$p_S$: unstable equilibrium. $p_S = \exp\{W_{-1}(-\hat{\lambda})\}$
$p_A$: undesired stable point
*S*: stable region of retransmission factor *q*.
$S_L$: absolute-stable region of retransmission factor *q*.
$S_A$: quasi-stable region of retransmission factor *q*.

## II. PRELIMINARY ANALYSIS

The buffered Aloha network resembles the statistical multiplexer in packet switching systems, as both of them can be considered as a system with multiple input queues contending for a single server. The main difference lies in their contention resolutions. When there is more than one packet request for the output in the statistical multiplexer, one packet will be selected randomly and dispatched to the output channel. In a buffered Aloha network, however, all packets contending for the same time slot will be dismissed.

As shown in Fig. 1, a buffered Aloha network can be modeled as a multi-queue-single-server system, in which each node is equipped with an infinite buffer and served by a common channel. The multi-queue-single-server system with *n* input queues can be characterized as a discrete-time Markov chain with a state space represented by the vector ($C_1$, $C_2$, ..., $C_n$), where $C_i$ is the queue length of node *i*. This multi-dimensional Markov chain is obviously intractable as the number of nodes *n* becomes too large.

In this paper, we adopt the decomposition approach in packet switching systems [26-28], and regard each node as an independent FIFO queue with identical Bernoulli arrival processes of rate *λ*. Both analytical and simulation results in packet switching systems have shown that this approximation is an effective approach with high accuracy to model the multi-queue systems.

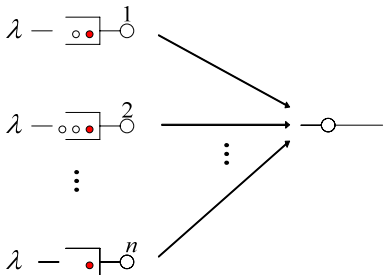

Fig. 1. An *n*-node buffered Aloha network can be modeled as an *n*-queue-single-server system.

### A. Modeling of Buffered Aloha

In a buffered Aloha network, only the HOL packets attend contention. Fig. 2 shows the state transition process of each individual HOL packet. The number of collisions experienced by a HOL packet is called the *phase* of the packet. Initially, a fresh HOL packet is in phase 0, and it moves to the next phase if it is involved in a collision. The contentions are resolved by backoff rescheduling algorithms. With *K-Exponential Backoff* protocol, a phase-*i* HOL packet has a transmission probability of $q^i$, *i*=0,1,...,*K*, where *q* is the *retransmission factor* and *K* is the *cutoff phase*.

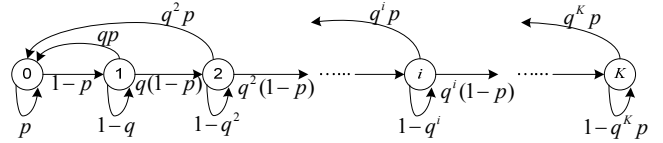

Fig. 2. State transition diagram of HOL packets.

The probability of success of each HOL packet, *p*, is assumed to be independent of the phase of the HOL packet. Intuitively, the chance that a HOL packet has a successful transmission should not vary with the number of collisions it has suffered. This assumption has been widely accepted and verified in various references [30].

Two particular backoff schemes are of special interest. *Geometric Retransmission* is a special case with cutoff phase *K*=1, that is, the retransmission probability is a constant *q* regardless of the number of collisions suffered. This collision resolution algorithm is the original version of slotted Aloha protocol that has been extensively investigated in [2-12, 19-25]. If the cutoff phase is unlimited, *K*=∞, then the protocol is simply called *Exponential Backoff*. For example, the binary exponential backoff (BEB) in previous studies [14-17] assumes *q*=1/2 and *K*=∞. Note that this assumption is slightly different from the one based on contention window [13, 18]. A detailed discussion will be presented in Part II of the paper series.

Let $f_0, f_1, ..., f_K$ represent the limiting probabilities of the Markov chain shown in Fig. 2. We have

$$f_i = f_{i-1}q^{i-1}(1-p) + f_i(1-q^i), \quad i=1,...,K\text{-}1, \quad (1)$$

and

$$f_K = f_{K-1}q^{K-1}(1-p) + f_K(1-q^K p). \quad (2)$$

It follows from (1-2) that

$$f_0 = 1 / \left( \frac{q}{p+q-1} - \left( \frac{q}{p+q-1} - \frac{1}{p} \right) \cdot \left( \frac{1-p}{q} \right)^K \right), \quad (3)$$

$$f_i = f_0 \left( \frac{1-p}{q} \right)^i, \quad i=1,...,K\text{-}1, \text{ and} \quad (4)$$

$$f_K = f_0 \left( \frac{1-p}{q} \right)^K / p. \quad (5)$$

Each fresh HOL packet will be in phase 0 for one time slot, and then be either transmitted or blocked. Hence, the service rate of each node's queue is $f_0$. Given the input rate *λ*, the offered load *ρ* per node can be obtained from (3):

$$\rho = \lambda / f_0 = \lambda \left( \frac{q}{p+q-1} - \left( \frac{q}{p+q-1} - \frac{1}{p} \right) \cdot \left( \frac{1-p}{q} \right)^K \right). \quad (6)$$

For Geometric Retransmission ($K=1$), the offered load $\rho$ is
$$\rho^{Geo} = \lambda(1-p+pq)/(pq). \quad (7)$$
In the case of Exponential Backoff ($K=\infty$), we have
$$\rho^{Exp} = \lambda q /(p+q-1). \quad (8)$$

The probability of success of each HOL packet, $p$, is determined by the contention level of the entire network. Theorem 1 presents its characteristic equation in steady-state.

**Theorem 1.** *For buffered Aloha with K-Exponential Backoff ($1 \leq K \leq \infty$), the steady-state probability of success p is given by*
$$p = \exp(-\hat{\lambda}/p), \quad (9)$$
*as the number of nodes $n \to \infty$, where $\hat{\lambda} = n\lambda$ is the aggregate input rate.*

Proof: Each node in the network must be in one of the following four states:
State 1: idle;
State 2: busy with a fresh HOL packet;
State 3: busy with an HOL packet in phase $i$ and retransmitting, $i=1, 2, \ldots, K$;
State 4: busy with an HOL packet in phase $i$ and not retransmitting, $i=1, 2, \ldots, K$.

We know that the probability of a node being busy is $\rho<1$, and the probability of an HOL packet being in phase $i$ given that the node is busy is $f_i$, $i=0,1,\ldots,K$. Therefore, the probability of the above four states are given by:
1) Pr{node is in State 1}=$1-\rho$;
2) Pr{node is in State 2}=$\rho f_0$;
3) Pr{node is in State 3}=$\rho f_i q^i$, $i=1,2,\ldots K$;
4) Pr{node is in State 4}=$\rho f_i (1-q^i)$, $i=1,2,\ldots K$.

When a node successfully transmits a packet, its $n-1$ interfering nodes must be either in State 1 or State 4. At steady-state, the probability of success $p$ of each HOL packet can be written as
$$p = (\Pr\{\text{node is in State 1}\} + \sum_{i=1}^{K} \Pr\{\text{node is in State 4, phase } i\})^{n-1}$$
$$= \left(1 - \rho + \sum_{i=1}^{K} \rho f_i (1-q^i)\right)^{n-1}. \quad (10)$$
Substituting (3-6) into (10), we have
$$p = \left(1 - \lambda \sum_{i=0}^{K-1}(1-p)^i - \lambda \frac{(1-p)^K}{p}\right)^{n-1} = (1-\lambda/p)^{n-1}. \quad (11)$$
It follows from (11) that (9) holds true as $n$ goes to infinity. □

Note that Theorem 1 is established under the assumption that the network is in steady-state. Besides, although the characteristic equation (9) is derived based on the independence assumption and an infinite population $n$, it is an effective approximation when $n$ is large enough, because the correlation among nodes' queues becomes weak as the network population $n$ grows [28]. The difference between (11) and (9) also rapidly decreases with the number of nodes $n$ increasing. The simulation results provided in Section V will corroborate that this approximation is quite accurate even for networks with a small population, for example $n=10$.

In a buffered network, the steady-state network throughput $\hat{\lambda}_{out}$ is equal to the aggregate input rate $\hat{\lambda}$. According to Theorem 1, the expected number of attempts per time slot, $G$, is given by
$$G = \hat{\lambda}_{out}/p = -\ln p. \quad (12)$$
It follows that the steady-state network throughput can be written as
$$\hat{\lambda}_{out} = \hat{\lambda} = G \exp(-G). \quad (13)$$
Note that in the original slotted Aloha model [3-4], the network throughput expression (13) was obtained by ignoring queueing at each node and considering the total number of attempts as a Poisson random variable with parameter $G$. In fact, the characteristic equation (9) (or equivalently, (13)) is solely determined by the contention of HOL packets of nodes, and hence holds true for Aloha networks with or without queueing taken into consideration.

In spite of the fact revealed in Theorem 1 that the steady-state probability of success is independent of the retransmission factor $q$ and the cutoff phase $K$, we will show that the selection of these backoff parameters, $q$ and $K$, is not arbitrary. Indeed, they are the keys to guarantee that stability can be achieved by various backoff protocols. This point will be fully elaborated in Sections III and IV.

*B. Bi-stable Property of Buffered Aloha*

The solutions of the fundamental characteristic equation (9) of probability of success $p$ can be represented by the Lambert W function defined by
$$W(z)e^{W(z)} = z, \quad (14)$$
which was first considered by J. Lambert around 1758, and later, studied by L. Euler [31]. The Lambert W function is a multivalued function. If $z$ is real and $-e^{-1}<z<0$, there are two possible real values of $W(z)$: the principal branch $W_0(z) \in [-1,\infty]$ and the other branch $W_{-1}(z) \in [-\infty,-1]$. Both branches are illustrated in Fig. 3.

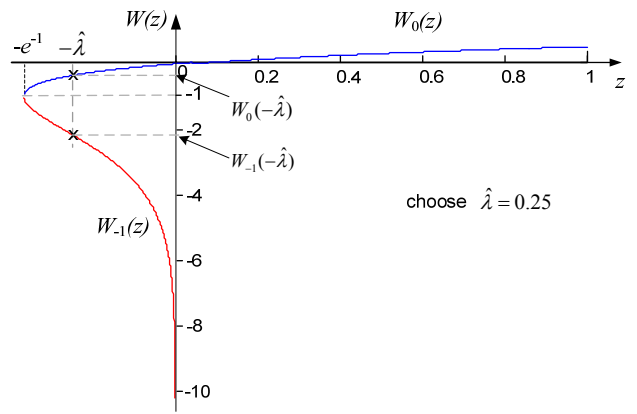

Fig. 3. The Lambert W function

The two non-zero solutions of (9) correspond, respectively, to the two branches of the Lambert W function, whose series expressions are given as follows:
1) $p_L = \exp(W_0(-\hat{\lambda}))$, and $W_0(z)$ has the following series expansion

$$W_0(z) = \sum_{i=1}^{\infty} \frac{(-i)^{i-1}}{i!} z^i = z - z^2 + \tfrac{3}{2}z^3 - \tfrac{8}{3}z^4 + \tfrac{125}{24}z^5 - \cdots \quad (15)$$

which can be derived by using the Lagrange inversion theorem. According to (15), $p_L$ can be further written as

$$p_L = \exp(W_0(-\hat{\lambda})) = 1 - \hat{\lambda} - \tfrac{1}{2}\hat{\lambda}^2 - \tfrac{2}{3}\hat{\lambda}^3 - \tfrac{63}{24}\hat{\lambda}^4 - \cdots . \quad (16)$$

2) $p_S = \exp(W_{-1}(-\hat{\lambda}))$, and $W_{-1}(z)$ has the following series expansion

$$W_{-1}(z) = \sum_{i=0}^{\infty} \mu_i x^i = -1 + x - \tfrac{1}{3}x^2 + \tfrac{11}{72}x^3 - \tfrac{43}{540}x^4 + \tfrac{769}{17280}x^5 - \cdots \quad (17)$$

in which $x = -\sqrt{2(ez+1)}$, and the coefficient $\mu_i$ is given in [31]. In general, we have $p_S \leq p_L$ and the equality holds for $p_S = p_L = e^{-1}$ when $\hat{\lambda} = e^{-1}$.

The previous stability analysis based on the drifts of the number of backlogged nodes has revealed that the slotted Aloha system has a desired stable point, an unstable equilibrium and an undesired stable point [7-8]. It was demonstrated that the system drifts to the desired stable point if and only if the attempt rate at any time slot is lower than the unstable equilibrium. Otherwise the system will drift to the undesired stable point. The drift analysis of bi-stable points and the unstable equilibrium is however described using the instantaneous number of backlogged nodes and is difficult to be applied for system control. In the following, we will demonstrate that the desired stable point and the unstable equilibrium are indeed given by $p_L$ and $p_S$, respectively.

Note that the exact dynamic trajectory of the instantaneous probability of success $p_t$ is difficult to be obtained. Here, we again resort to approximation. The state distribution of each single HOL packet at time slot $t$ is approximated by the steady-state distribution with transition characteristics at time slot $t$. The accuracy of this approximation can be easily verified by simulations and a similar approach has been adopted in [32].

Specifically, suppose that at time slot $t$, the probability of an HOL packet being in phase $i$ given that the node is busy is $f_{i,t}$, $i=0, 1, \ldots, K$. We approximate $f_{i,t}$ by

$$f_{i,t} \approx f_i(p_t), \quad (18)$$

where $f_i$ is the stationary state distribution given in (3-5), $i=0, 1, \ldots, K$. By following a similar derivation as shown in Theorem 1, the probability of success at time slot $t+1$, $p_{t+1}$, can be written as

$$p_{t+1} = \left(1 - \rho_t + \sum_{i=1}^{K} \rho_t f_{i,t}(1-q^i)\right)^{n-1} \quad (19)$$

where $\rho_t$ is the offered load at time slot $t$ given by

$$\rho_t = \lambda / f_{0,t} < 1. \quad (20)$$

Combining (18-20), finally we have

$$p_{t+1} \approx (1 - \lambda / p_t)^{n-1} \overset{\text{with a large } n}{\approx} \exp(-\hat{\lambda} / p_t). \quad (21)$$

Fig. 4 presents the approximate dynamic trajectory of the instantaneous probability of success $p_t$. It can be clearly seen from Fig. 4 that $\exp(-\hat{\lambda}/p_t)$ is a contraction mapping on the interval $(p_S, 1]$, and has a unique fixed point $p_L$. As a result, the probability of success will ultimately converge to $p_L$ if at any time slot $p_t > p_S$, indicating that $p_L$ is the desired stable point.

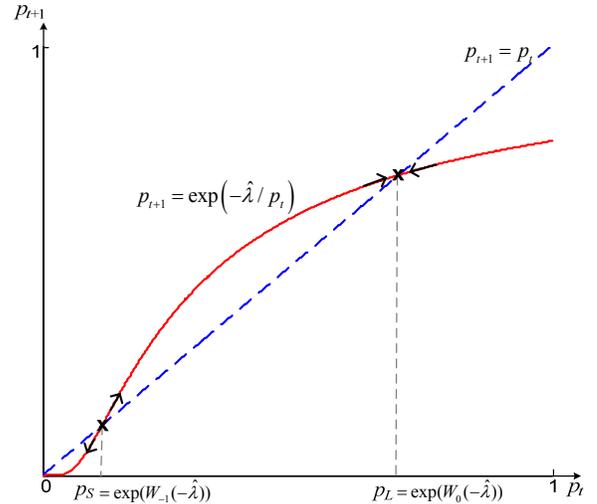

Fig. 4. Dynamic trajectory of the instantaneous probability of success $p_t$ based on approximation (18).

On the other hand, if the probability of success $p_t$ drops below $p_S$, it will be departing from $p_S$ because $p_{t+1} = \exp(-\hat{\lambda}/p_t) < p_t < p_S$. It can then be concluded that $p_S$ is the unstable equilibrium. Moreover, as the probability of success $p_t$ becomes smaller and smaller, all the nodes will eventually become busy and the network is saturated, in which case $\rho_t$ becomes 1. As a consequence, the probability of success $p_t$ will not be governed by (21). Instead of zero, $p_t$ will converge to the undesired stable point $p_A$ that depends on the backoff protocols. A detailed discussion on the undesired stable point will be presented in Section IV.

As shown in Fig. 5, the system dynamics described above can be also demonstrated in terms of the attempt rate $G_t$ at time slot $t$: If $G_t$ is to the left of the unstable equilibrium point $-W_{-1}(-\hat{\lambda})$, then $G_t$ will converge to the desired stable point $-W_0(-\hat{\lambda})$ as $t\to\infty$. Otherwise it will drift to the undesired stable point.

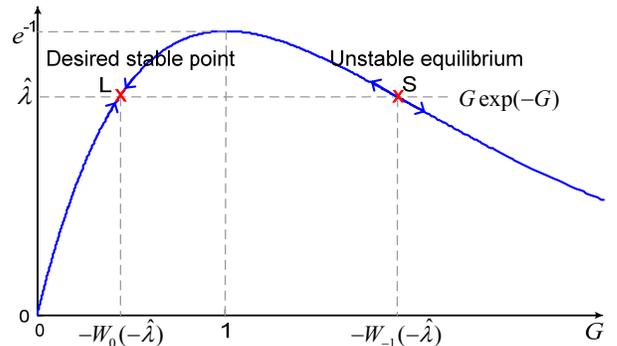

Fig. 5. System drift in terms of attempt rate $G$

## III. ABSOLUTE-STABLE REGION

We have demonstrated in Section II that the steady-state network throughput of $\hat{\lambda}$ can be achieved when the network operates at the desired stable point $p_L$. In this section, we will further investigate the conditions that guarantee the convergence of probability of success to the desired stable point $p_L$.

It is shown in Section II. B that the convergence of probability of success to $p_L$ requires that

$$p_t \geq p_S = \exp\{W_{-1}(-\hat{\lambda})\}, \quad (22)$$

or equivalently, the instantaneous attempt rate $G_t$ satisfies

$$G_t \leq -\ln p_S = -W_{-1}(-\hat{\lambda}) \quad (23)$$

at any time slot $t$. In the next theorem, we will show that this constraint imposes an upper bound on the retransmission factor $q$.

**Theorem 2.** For buffered Aloha with K-Exponential Backoff ($1 \leq K \leq \infty$), if

$$q \leq q_u = -\ln p_S / n = -W_{-1}(-\hat{\lambda}) / n, \quad (24)$$

then at any time slot $t$, $G_t \leq -\ln p_S$.

Proof: Suppose that there are totally $n_b$ backlogged HOL packets at time slot $t$, with $n_i$ packets in phase $i$, $i=1,\ldots,K$. The attempt rate $G_t$ is then given by

$$G_t = (n-n_b)\lambda + \sum_{i=1}^{K} n_i q^i \leq (n-n_b)\lambda + n_b q \quad (25)$$

where the right side of (25) is the attempt rate corresponding to the state that $n_b$ backlogged HOL packets are all in phase 1. We consider the following two cases:

1) If retransmission factor $q \leq \lambda$, the attempt rate $G_t$ is bounded by

$$G_t \leq \hat{\lambda}. \quad (26)$$

We know from Section II. B that

$$-\ln p_S = -W_{-1}(-\hat{\lambda}) > \hat{\lambda}. \quad (27)$$

By combining (26) and (27), we have

$$G_t \leq -\ln p_S. \quad (28)$$

2) If retransmission factor $q \geq \lambda$, the attempt rate $G_t$ is bounded by

$$G_t \leq nq \leq nq_u = -\ln p_S. \quad (29)$$

Hence, the theorem is established by combining (28) and (29). □

Another vital criterion imposed on the range of retransmission factor $q$ is that the offered load $\rho$ of each input queue given by (6) must be no larger than 1 to ensure the throughput stability. The lower bound of the retransmission factor $q$ is specified in the following theorem.

**Theorem 3.** For buffered Aloha with K-Exponential Backoff ($1 \leq K \leq \infty$), suppose the probability of success $p = p_L$, then $\rho \leq 1$ iff $q \geq q_l$, where the lower bound $q_l$ is the root of the following equation:

$$\frac{q}{p_L+q-1} - \left(\frac{q}{p_L+q-1} - \frac{1}{p_L}\right) \cdot \left(\frac{1-p_L}{q}\right)^K = 1/\lambda. \quad (30)$$

In particular, the lower bound $q_l$ for Geometric Retransmission ($K=1$) is

$$q_l^{Geo} = \frac{\lambda(1-p_L)}{p_L(1-\lambda)} = \frac{\hat{\lambda}(1-p_L)}{p_L(n-\hat{\lambda})}, \quad (31)$$

and for Exponential Backoff ($K=\infty$), we have

$$q_l^{Exp} = \frac{1-p_L}{1-\lambda} = \frac{1-p_L}{1-\hat{\lambda}/n}. \quad (32)$$

Proof: As the probability of success $p=p_L$, it is easy to show from (6) that the offered load $\rho$ monotonically increases with $(1-p_L)/q$, and hence $\rho$ is a monotonic decreasing function of $q$.

It follows that the minimum retransmission factor $q$ corresponds to the maximum offered load $\rho$. Thus, the lower bound of $q$ is the root of $\rho=1$, which transforms (6) into (30). □

Note that the retransmission factor $q$ should be strictly larger than $q_l$ if delay stability, i.e., a finite mean queue length, is required. Besides, according to Theorem 2, $q$ should not exceed $q_u$ to guarantee that the probability of success converges to $p_L$. The offered load $\rho$ versus the retransmission factor $q$ is plotted in Fig. 6 under different values of cutoff phase $K$. It shows that the offered load $\rho$ is a monotonic decreasing function of $q$ for any given $K$. Fig. 6 also indicates that a larger cutoff phase $K$ leads to a higher offered load $\rho$, which incurs a larger access delay, for any given retransmission factor $q$.

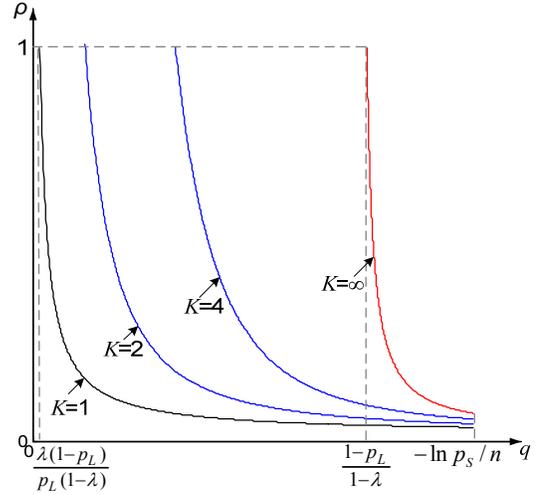

Fig. 6. Tradeoff between offered load $\rho$ and retransmission factor $q$.

Based on Theorems 2 and 3, we define the *absolute-stable region* of retransmission factor $q$ as

$$S_L = [q_l, q_u], \quad (33)$$

where $q_l$ and $q_u$ are given in (30) and (24), respectively. If the retransmission factor $q$ is chosen from the absolute-stable region, $q \in S_L$, the network will stabilize at the desired stable point $p_L$ for sure. The absolute-stable region $S_L$ is dependent on the aggregate input rate $\hat{\lambda}$ and the total number of nodes $n$. We can see from (24) and (31-32) that the region $S_L$ may become an empty set if either the aggregate input rate $\hat{\lambda}$ is too large or the number of nodes $n$ goes to infinity.

Define $\hat{\lambda}_{\max\_S_L}$ as the maximum stable throughput that the network can achieve when the retransmission factor $q$ is selected from the absolute-stable region $S_L$:

$$\hat{\lambda}_{\max\_S_L} = \sup_{S_L} \hat{\lambda}. \quad (34)$$

Given that $q_l$ and $q_u$ are monotonic increasing and decreasing functions, respectively, of the aggregate input rate $\hat{\lambda}$, the maximum stable throughput $\hat{\lambda}_{\max\_S_L}$ should be the single root of

$$q_l(\hat{\lambda}) - q_u(\hat{\lambda}) = 0.$$

We will take Geometric Retransmission ($K=1$) and Exponential Backoff ($K=\infty$) as two examples to demonstrate the above results in the following subsections.

*A. Absolute-Stable Region of Geometric Retransmission (K=1)*

The absolute-stable region $S_L^{Geo}$ of Geometric Retransmission can be obtained by combining (24) and (31):

$$S_L^{Geo} = \left[ q_l^{Geo}, \ q_u^{Geo} \right] = \left[ \frac{\hat{\lambda}(1-p_L)}{p_L(n-\hat{\lambda})}, \ -\frac{\ln p_S}{n} \right]. \quad (35)$$

The difference between the upper bound $q_u^{Geo}$ and the lower bound $q_l^{Geo}$ is given by

$$q_u^{Geo}(\hat{\lambda}) - q_l^{Geo}(\hat{\lambda}) \approx \tfrac{1}{n}\left( -\ln p_S - \hat{\lambda}(1-p_L)/p_L \right), \quad (36)$$

which is a monotonic decreasing function of $\hat{\lambda}$. When $\hat{\lambda}=e^{-1}$, the difference $e^{-1}/n$ approaches zero with a large number of nodes $n$. Therefore, we can conclude that the maximum stable throughput with $q \in S_L^{Geo}$ is given by

$$\hat{\lambda}_{\max\_S_L}^{Geo} \approx e^{-1}, \quad (37)$$

and the corresponding retransmission factor $q=1/n$.

As shown in Fig. 7, the absolute-stable region $S_L^{Geo}$ becomes narrower and narrower as the aggregate input rate $\hat{\lambda}$ increases. It eventually shrinks to $q=1/n$, with which the aggregate input rate $\hat{\lambda}=\hat{\lambda}_{\max\_S_L}^{Geo}$.

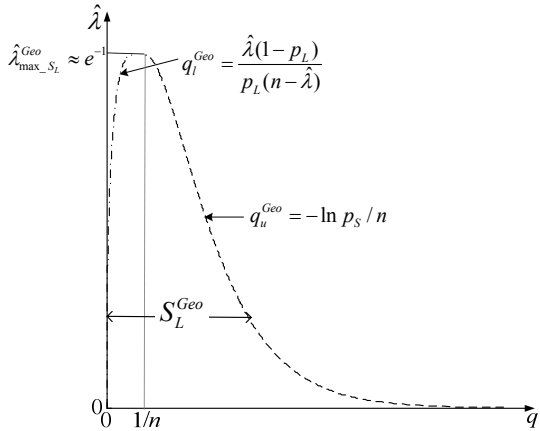

Fig. 7. Absolute-stable region and the corresponding maximum stable throughput of Geometric Retransmission (*K*=1).

Both the upper bound $q_u^{Geo}$ and the lower bound $q_l^{Geo}$ will approach zero, and the absolute-stable region of Geometric Retransmission will vanish, $S_L^{Geo}=\varnothing$, when the number of nodes $n \to \infty$. It indicates that the network cannot be stabilized at the desired stable point $p_L$ when the number of nodes $n$ is infinite.

*B. Absolute-Stable Region of Exponential Backoff (K=∞)*

The absolute-stable region $S_L^{Exp}$ of Exponential Backoff can be obtained by combining (24) and (32):

$$S_L^{Exp} = \left[ q_l^{Exp}, \ q_u^{Exp} \right] = \left[ \frac{1-p_L}{1-\hat{\lambda}/n}, \ -\frac{\ln p_S}{n} \right]. \quad (38)$$

It is known from (16) that $1-\hat{\lambda}$ is a good approximation for $p_L$. Therefore, the lower bound $q_l^{Exp}$ can be approximated by

$$q_l^{Exp} = \frac{1-p_L}{1-\hat{\lambda}/n} \approx \frac{\hat{\lambda}}{1-\hat{\lambda}/n} \approx \hat{\lambda} \quad (39)$$

when the number of nodes $n$ is large. It follows from (24) and (39) that the maximum stable throughput $\hat{\lambda}_{\max\_S_L}^{Exp}$ is given by:

$$\hat{\lambda}_{\max\_S_L}^{Exp} \approx \ln n / n, \quad (40)$$

and the corresponding retransmission factor $q \approx \hat{\lambda}_{\max\_S_L}^{Exp} \approx \ln n/n$.

The absolute-stable region of Exponential Backoff $S_L^{Exp}$ is depicted in Fig. 8. In contrast to Geometric Retransmission, here the maximum stable throughput $\hat{\lambda}_{\max\_S_L}^{Exp} \approx \ln n/n$ is much lower than $e^{-1}$ when the number of nodes $n$ is large, and decreases rapidly with $n$ increasing. Again, the absolute-stable region $S_L^{Exp}$ becomes an empty set as the number of nodes $n \to \infty$.

On the other hand, with a finite number of nodes $n$, the network can be stabilized at the desired stable point $p_L$ if the aggregate input rate $\hat{\lambda}$ is lower than $\hat{\lambda}_{\max\_S_L}^{Exp} \approx \ln n/n$. Note that it was demonstrated in [15-17] that the network with binary exponential backoff (BEB) is stable if the arrival rate is smaller than $\lambda^*(n)$, where $\lambda^*(n)$ is a decreasing function of the number of nodes $n$. In Fig. 8, it is easy to see that if the aggregate input rate $\hat{\lambda}$ is lower than $\tfrac{1}{2}ne^{-n/2}$, then the retransmission factor $q=1/2$ is included in the absolute-stable region $S_L^{Exp}$. In other words, the network with BEB can be stabilized at the desired stable point $p_L$ if the aggregate input rate $\hat{\lambda} \le \lambda^*(n) = \tfrac{1}{2}ne^{-n/2}$.

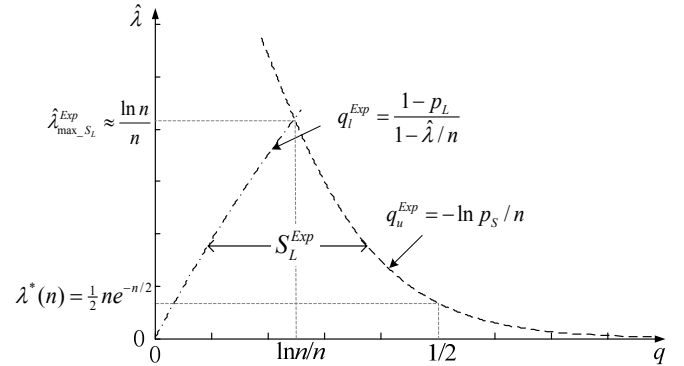

Fig. 8. Absolute-stable region and the corresponding maximum stable throughput of Exponential Backoff (*K*=∞).

*C. Absolute-Stable Region of K-Exponential Backoff (1<K<∞)*

For the general *K*-Exponential Backoff with $1<K<\infty$, it is quite difficult to obtain an explicit expression for the lower bound $q_l$. In Appendix I, we show that with a large number of nodes $n$, the absolute-stable region $S_L^{K-Exp}$ of *K*-Exponential Backoff with $1<K<\infty$ is approximately given by:

$$S_L^{K-Exp} = \left[ q_l^{K-Exp}, \ q_u^{K-Exp} \right] \approx \left[ \frac{1-p_L}{\sqrt[K]{np_L/\hat{\lambda}}}, \ -\ln p_S/n \right], \quad (41)$$

and the corresponding maximum stable throughput is

$$\hat{\lambda}_{\max\_S_L}^{K-Exp} \approx \ln n^{1-1/K} / n^{1-1/K}, \quad (42)$$

which declines with an increasing cutoff phase *K*, and is consistent with (40) when *K*=∞. We can see from (41) that both $q_l^{K-Exp}$ and $q_u^{K-Exp}$ approach zero as the number of nodes $n$ increases, indicating that *K*-Exponential Backoff with $1<K<\infty$ cannot be stabilized at the desired stable point $p_L$ either, when the number of nodes $n$ is infinite.

The sum and substance of this section is that a network with K-Exponential Backoff can be stabilized at the desired stable point $p_L$ for sure if the retransmission factor $q$ is selected from the absolute-stable region $S_L$. Both throughput and delay stabilities are achievable at the desired stable point $p_L$.

With an increase of the number of nodes $n$, however, $S_L$ will rapidly shrink and vanish as $n\to\infty$, indicating that the network with K-Exponential Backoff cannot be stabilized at $p_L$ with an infinite number of nodes $n$. Moreover, the corresponding maximum stable throughput could be diminished by increasing the cutoff phase $K$. With Exponential Backoff ($K=\infty$), the maximum stable throughput will quickly become zero as the number of nodes $n$ grows.

Note that operating at the desired stable point $p_L$ may not be a necessary condition for the network to achieve throughput stability. In the next section, we will demonstrate that a network with Exponential Backoff can still have a stable throughput even when it shifts to the undesired stable point $p_A$. The maximum stable throughput of $e^{-1}$ can be achieved, but at a cost of severely deteriorated delay performance.

## IV. QUASI-STABLE REGION

This section is devoted to the analysis of network behavior when the network shifts to *undesired* stable point $p_A$.

It is shown in Section II. B that the instantaneous probability of success $p_t$ will monotonically decrease if $p_t$ drops below the unstable equilibrium $p_S$. Consequently, the service rate $f_{0,t}$ of each queue will become smaller and smaller and eventually drop below the input rate $\lambda$. In this case, the network becomes saturated and all nodes in the network will be busy with the offered load $\rho_t=1$. According to (19), if all $n-1$ interfering nodes are busy, the probability of success at time slot $t+1$ can be written as:

$$p_{t+1} = \left(\sum_{i=1}^{K} f_{i,t}(1-q^i)\right)^{n-1} \overset{\text{with a large } n}{\approx} \exp(-nf_{0,t}/p_t). \quad (43)$$

Substituting (18) and (3) into (43), we have

$$p_{t+1} = \exp\{-n/g(p_t)\}, \quad (44)$$

where

$$g(p_t) = \frac{p_t q}{p_t+q-1} - \left(\frac{p_t q}{p_t+q-1}-1\right)\cdot\left(\frac{1-p_t}{q}\right)^K \quad (45)$$

is a monotonic decreasing function of $p_t$. The single root of equation $p=\exp\{-n/g(p)\}$ is shown in Fig. 9.

It is straightforward to show that:
i) When $p_{t-1} < p_t$, according to $g(p_{t-1}) > g(p_t)$ we have $p_t > p_{t+1}$;
ii) When $p_{t-1} > p_t$, according to $g(p_{t-1}) < g(p_t)$ we have $p_t < p_{t+1}$.
Therefore, $p_t$ converges to the unique fixed point $p_A$ as $t\to\infty$, which is the root of the following equation:

$$p = \exp\left\{-n/\left(\frac{pq}{p+q-1}-\left(\frac{pq}{p+q-1}-1\right)\cdot\left(\frac{1-p}{q}\right)^K\right)\right\}. \quad (46)$$

It can be clearly seen from (46) that the undesired stable point $p_A$ is no longer determined by the traffic input rate $\hat{\lambda}$. Instead, it becomes a function of backoff parameters such as the retransmission factor $q$ and the cutoff phase $K$. This indicates that $p_A$ may vary under different backoff protocols.

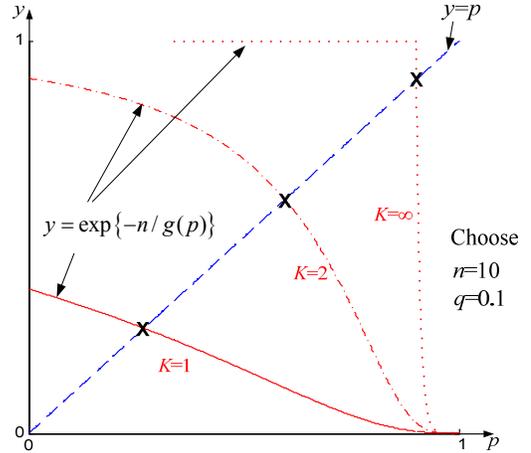

Fig. 9. Root of Equation $p=\exp\{-n/g(p)\}$.

From (3) and (46), we can see that when the network operates at the undesired stable point $p_A$, the aggregate service rate is given by

$$nf_0 = -p_A \ln p_A. \quad (47)$$

To achieve a stable network throughput of $\hat{\lambda}$, the aggregate service rate should be no less than the aggregate input rate, that is,

$$nf_0 \geq \hat{\lambda}. \quad (48)$$

According to (47-48), when the network operates at the undesired stable point $p_A$, the corresponding stable region of retransmission factor $q$ can be defined as follows:

$$S_A = \{q \mid p_S \leq p_A \leq p_L\}. \quad (49)$$

Outside this region, $q \notin S_A$, the network will become unstable and the network throughput is given by

$$\hat{\lambda}_{out} = -p_A \ln p_A < \hat{\lambda} \quad (50)$$

as shown in Fig. 10.

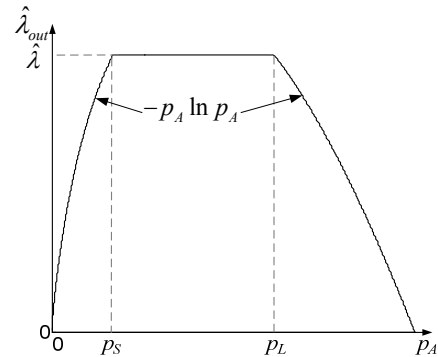

Fig. 10. Network throughput when the network operates at the undesired stable point $p_A$.

In contrast to the desired stable point $p_L$, at which both throughput and delay stabilities are achievable, networks operating at $p_A$ may become unstable. In the following subsections, we will discuss the stable regions of Geometric Retransmission ($K=1$) and Exponential Backoff ($K=\infty$) in detail.

## A. Stable Region of Geometric Retransmission

When the cutoff phase $K=1$, we know from (46) that the undesired stable point of Geometric Retransmission, $p_A^{Geo}$, should satisfy the following equation:

$$p = \exp(-nq/(1-p+pq)). \quad (51)$$

For large $n$, $p_A^{Geo}$ can be approximately given by

$$p_A^{Geo} \approx \exp\{-nq\}. \quad (52)$$

We have demonstrated in Section III that the network will operate at the desired stable point $p_L$ if the retransmission factor $q$ is selected from the absolute-stable region $q \in S_L^{Geo}$. It is easy to show that with $q \notin S_L^{Geo}$, the undesired stable point $p_A^{Geo}$ will be either higher than $p_L$, or lower than $p_S$:

1) If $q > q_u^{Geo} = -\ln p_S / n$, then (52) implies $p_A^{Geo} < p_S$;
2) If $q < q_l^{Geo} = \frac{\hat{\lambda}(1-p_L)}{p_L(n-\hat{\lambda})}$, then according to (52) we have

$$p_A^{Geo} \approx \exp\{-\hat{\lambda}(1-p_L)/p_L\} > \exp\{-\hat{\lambda}/p_L\} = p_L. \quad (53)$$

Therefore, we can conclude from (49) that $S_A^{Geo} = \emptyset$.

*The complete stable region of Geometric Retransmission is given by*

$$S^{Geo} = S_L^{Geo} \cup S_A^{Geo} = S_L^{Geo}, \quad (54)$$

*and the maximum stable throughput is*

$$\hat{\lambda}_{max\_S}^{Geo} = \hat{\lambda}_{max\_S_L}^{Geo} = e^{-1}, \quad (55)$$

*with the corresponding retransmission factor $q=1/n$.*

Figs. 11 and 12 show the probability of success $p$ and the network throughput of Geometric Retransmission under different values of retransmission factor $q$, respectively. It can be clearly seen that with Geometric Retransmission, the network is stable iff $q \in S_L^{Geo}$. The probability of success $p$ will converge to the desired stable point $p_L$, at which both throughput and delay stabilities can be achieved.

With an increase of the number of nodes $n$, the stable region $S^{Geo}$ will rapidly shrink and finally vanish as $n \to \infty$. Therefore, we can conclude that the network with Geometric Retransmission is unstable if the number of nodes is infinite. This is consistent with the previous studies [5-6] that the slotted Aloha network with Geometric Retransmission is inherently *unstable* with an infinite population.

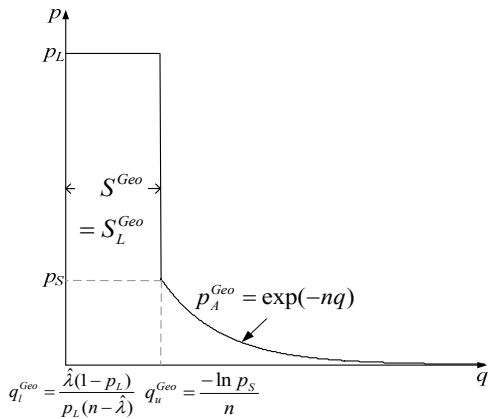

Fig. 11. Probability of success $p$ versus retransmission factor $q$ in the Geometric Retransmission case.

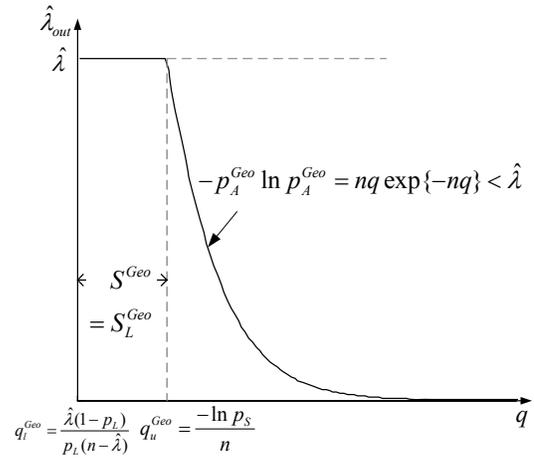

Fig. 12. Network throughput versus retransmission factor $q$ in the Geometric Retransmission case.

In fact, with an increase of the number of nodes $n$, the contention grows, and backlogged HOL packets have to back off to larger phases so as to reduce the attempt rate. For Geometric Retransmission, the contention level cannot be alleviated because all backlogged HOL packets will stay in phase 1, and the corresponding attempt rate $G=nq$ will keep growing with $n$ under any given retransmission factor $q$. As a result, the network will eventually become unstable as the number of nodes $n$ goes to infinity.

For Exponential Backoff, however, there is no limit on the phases of HOL packets. Backlogged HOL packets can always back off to deeper phases to alleviate contentions, and to make the attempt rate arbitrarily small until the network is stabilized. In the next subsection, we will show that a stable throughput can be achieved with Exponential Backoff even when the number of nodes $n \to \infty$.

## B. Stable Region of Exponential Backoff

When the cutoff phase $K=\infty$, we know from (46) that the undesired stable point of Exponential Backoff, $p_A^{Exp}$, should satisfy the following equation:

$$p = \exp\left(-n \cdot \frac{q-(1-p)}{pq}\right). \quad (56)$$

For large $n$, the derivation presented in Appendix II shows that $p_A^{Exp}$ can be approximately given by

$$p_A^{Exp} \approx 1-q. \quad (57)$$

It is shown in Section III that the absolute-stable region of Exponential Backoff will quickly become empty, $S_L^{Exp} = \emptyset$, when the number of nodes $n$ is large. Outside the absolute-stable region, when the network operates at the undesired stable point $p_A^{Exp}$, it can be easily derived from (49) and (57) that $S_A^{Exp}$ is given by

$$S_A^{Exp} = [1-p_L, 1-p_S]. \quad (58)$$

It is plain to see from (58) that $S_A^{Exp}$ shrinks as the aggregate input rate $\hat{\lambda}$ increases, and eventually becomes a single point $\{1-e^{-1}\}$ when $\hat{\lambda}$ reaches the maximum stable throughput $\hat{\lambda}_{max\_S_A}^{Exp} = e^{-1}$.

*For large n, the complete stable region of Exponential Backoff is given by*

$$S^{Exp} = S_L^{Exp} \cup S_A^{Exp} = S_A^{Exp}, \quad (59)$$

*and the maximum stable throughput is*

$$\hat{\lambda}_{max\_S}^{Exp} = \hat{\lambda}_{max\_S_A}^{Exp} = e^{-1}, \quad (60)$$

*with the corresponding retransmission factor $q=1-e^{-1}$.*

In contrast to Geometric Retransmission, throughput stability is achievable in the network with Exponential Backoff even when it operates at the undesired stable point $p_A^{Exp}$. As shown in Fig. 13, when the traffic level is low, the network can operate at the desired stable point $p_L$ for sure with $q \in S_L^{Exp}$. With an increase of aggregate input rate $\hat{\lambda}$, however, the absolute-stable region $S_L^{Exp}$ will become empty. Nevertheless, a stable throughput can still be achieved at the undesired stable point $p_A^{Exp}$ with $q \in S_A^{Exp}$.

It can be also seen from Fig. 13 that $q=1/2$ is included in the stable region $S^{Exp}$ if the aggregate input rate $\hat{\lambda} \leq \frac{1}{2}\ln 2$. In other words, a network with BEB ($q=1/2$) can be stabilized if the traffic input rate $\hat{\lambda} \leq \frac{1}{2}\ln 2$, which is consistent with the result proved in [29] that the network with BEB is unstable if $\hat{\lambda}>\ln 2$.

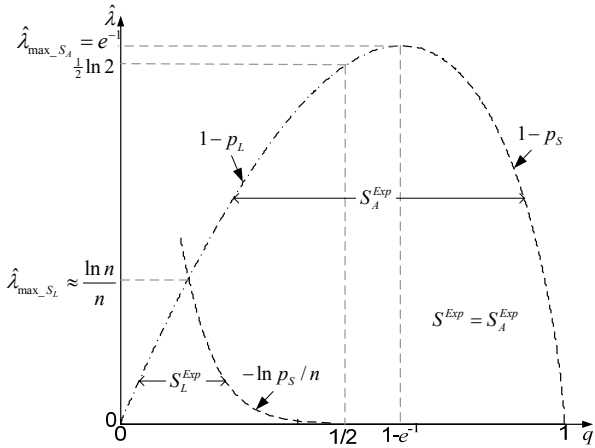

Fig. 13. Stable region and maximum stable throughput of Exponential Backoff.

Fig. 14 further presents the network throughput of Exponential Backoff under different values of retransmission factor *q*. Note that the stable region $S^{Exp}=[1-p_L, 1-p_S]$ does not vary with the number of nodes *n*, indicating that the network with Exponential Backoff is highly robust. It can be stabilized even when $n\to\infty$.

Despite its robustness, the network with Exponential Backoff may suffer from severe delay jitter when operating at the undesired stable point $p_A^{Exp}$. In particular, when the network becomes saturated, nodes would have to back off to deeper phases with extremely small retransmission probabilities. As a result, once a node tries to retransmit and succeeds, it is very likely that this node will dominate the channel for a fairly long period of time and produce a continuous stream of packets until it is interrupted by the retransmission requests initiated by other backlogged nodes. This "capture phenomenon" occurring when the network becomes saturated has been described in [33-35].

It will be demonstrated in Part II of the paper series that the mean queueing delay of input packets will grow unboundedly when the Exponential Backoff system operates at the undesired stable point $p_A$. Despite a stable throughput, the network is indeed quasi-stable with $q \in S_A^{Exp}$. The stable region $S_A^{Exp}$ is therefore referred to as *quasi-stable region* of Exponential Backoff.

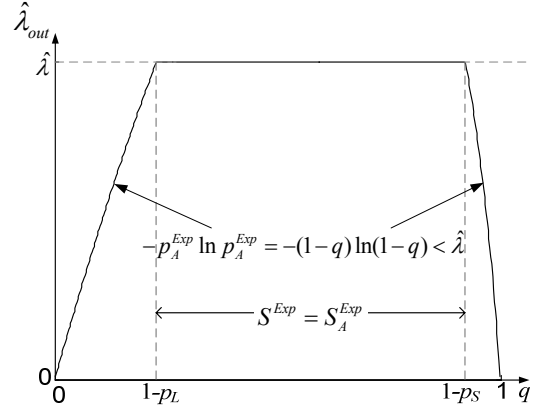

Fig. 14. Network throughput versus retransmission factor *q* in the Exponential Backoff case.

For *K*-Exponential Backoff with $1<K<\infty$, an explicit expression for undesired stable point $p_A$ with a general cutoff phase *K* is rather difficult to be obtained. In Appendix III, we provide the proof of the following distinguishing characteristics of $p_A$ with a finite cutoff phase $1<K<\infty$:

1) $p_A$ *is monotonic increasing with respect to cutoff phase K;*
2) *For any given retransmission factor q, $p_A \to 0$ as $n\to\infty$.*

We have shown in Section III that the absolute-stable region becomes empty for any cutoff phase *K* as $n\to\infty$. The above characteristics further indicate that although $p_A$ can be improved with a larger cutoff phase *K*, the *K*-Exponential Backoff with $1<K<\infty$ remains *unstable* if the number of nodes *n* is infinite.

## V. SIMULATION RESULTS

In this section, we will provide the simulation results to verify the preceding theoretical analysis. We first consider a small network (*n*=10 nodes) with light traffic (the aggregate input rate $\hat{\lambda}$=0.1). Fig. 15 presents the curves of the offered load $\rho$ versus the retransmission factor *q* under different values of cutoff phase *K*. The simulation results well agree with (6) in the absolute-stable region of *q*. For the sake of clarity, in the following figures we only provide the corresponding curves of Geometric Retransmission (*K*=1) and Exponential Backoff (*K*=∞).

Simulation results shown in Fig. 16 demonstrate that with *q* selected from the absolute-stable region $S_L$, the probability of success *p* will converge to the desired stable point $p_L$, at which a stable throughput can be achieved. The desired stable point $p_L$ is solely determined by the aggregate input rate $\hat{\lambda}$, and is invariant with respect to the system parameters such as retransmission factor *q* and cutoff phase *K*.

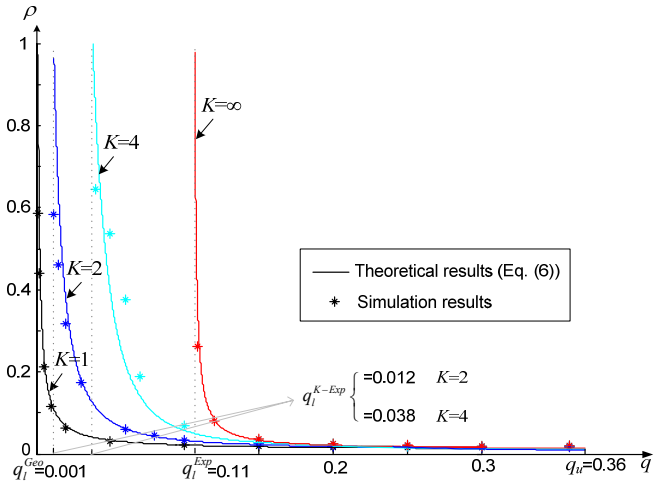

Fig. 15. Offered load $\rho$ versus retransmission factor $q$ within the absolute-stable region $S_L=[q_l, q_u]$. $n=10$ and $\hat{\lambda} = 0.1$.

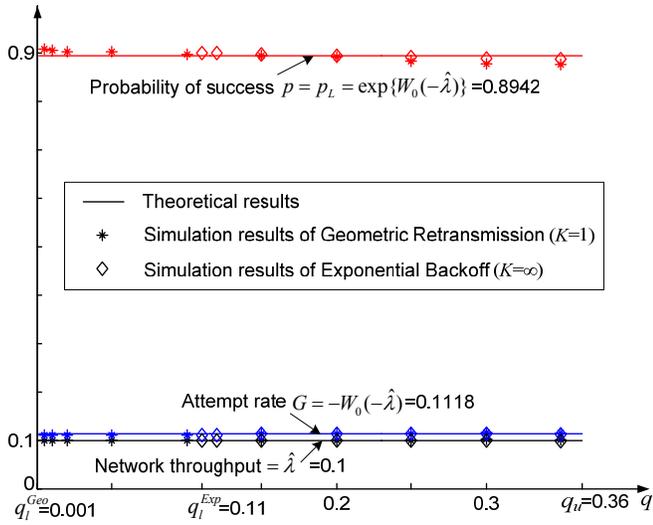

Fig. 16. Probability of success $p$, attempt rate $G$ and network throughput versus retransmission factor $q$ within the absolute-stable region $S_L=[q_l, q_u]$. $n=10$ and $\hat{\lambda} = 0.1$.

The absolute-stable region $S_L$ will be rapidly diminishing with an increase of either the number of nodes $n$, or the aggregate input rate $\hat{\lambda}$. For example, when $n=50$ and $\hat{\lambda}=0.3$, the absolute-stable region of Geometric Retransmission $S_L^{Geo}$ becomes [0.0038, 0.0356] according to (35). We have demonstrated in Section IV. A that outside the absolute-stable region $S_L^{Geo}$, the probability of success will converge to the undesired stable point $p_A^{Geo} \approx \exp(-nq)$, which sharply decays with the retransmission factor $q$. For Exponential Backoff, the absolute-stable region $S_L^{Exp}$ becomes empty with $n=50$ and $\hat{\lambda}=0.3$ according to (38). The probability of success will converge to the undesired stable point $p_A^{Geo} \approx 1-q$, as we have demonstrated in Section IV. B. Both cases have been verified in Fig. 17.

Fig. 18 shows the corresponding throughput performance. With Geometric Retransmission, the network throughput quickly approaches zero when the retransmission factor $q$ exceeds the stable region $S^{Geo}= S_L^{Geo}$. As for Exponential Backoff, a network throughput of $\hat{\lambda}_{out}=\hat{\lambda}=0.3$ is achievable if the retransmission factor $q$ is chosen from the stable region $S^{Exp} = S_A^{Exp} = [0.387, 0.8316]$. Outside the stable region, the network becomes unstable, and is less predictable due to the non-stationary queueing behavior of each individual node.

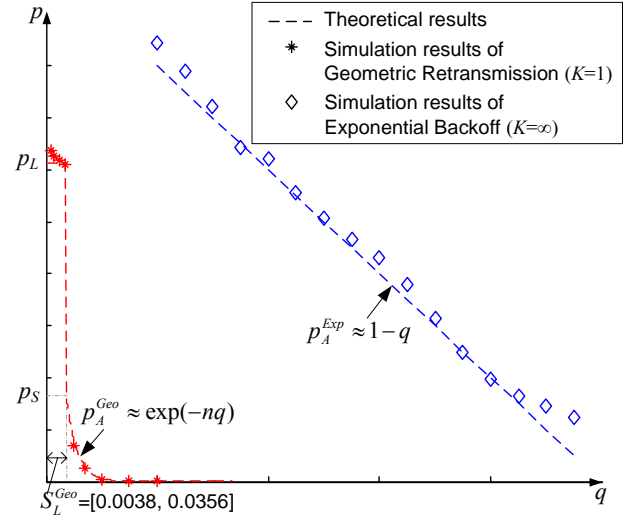

Fig. 17. Probability of success $p$ versus retransmission factor $q$ with $n=50$ and $\hat{\lambda} = 0.3$.

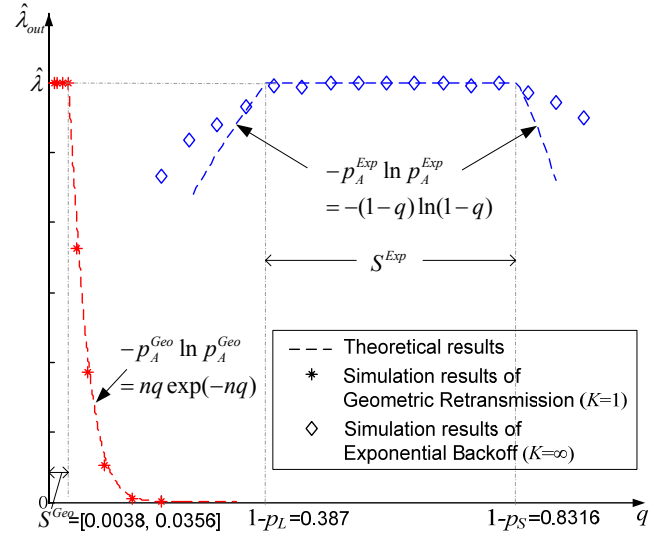

Fig. 18. Network throughput versus retransmission factor $q$ with $n=50$ and $\hat{\lambda} = 0.3$.

The simulation results presented in this section corroborate that throughput stability can be achieved in buffered Aloha networks with *K*-Exponential Backoff if the retransmission factor $q$ is properly selected from the corresponding stable region. Exponential Backoff has a much larger stable region than Geometric Retransmission, implying that better system robustness is provided by Exponential Backoff. Nevertheless, as we will further demonstrate in Part II of the paper series, Exponential Backoff networks suffer from severely deteriorated delay performance with a high aggregate input rate. The network will become quasi-stable when operating at the undesired stable point.

## VI. CONCLUSIONS

In this paper, a unified approach is developed to characterize the stable region and throughput of a buffered Aloha network with *K*-Exponential Backoff collision resolution algorithms. We demonstrate that a network with *K*-Exponential Backoff can be stabilized at the desired stable point $p_L$ if the retransmission factor $q$ is selected from the absolute-stable region $S_L$. Both throughput and delay stabilities are achievable at the desired stable point $p_L$. For large cutoff phase $K$, however, the absolute-stable region will vanish with a large network population $n$, or a high traffic input rate $\hat{\lambda}$. The analysis of the undesired stable point $p_A$ further reveals that a stable throughput can still be achieved at $p_A$ provided that the cutoff phase $K$ is large enough. We derive the stable region of Exponential Backoff ($K=\infty$) and show that it does not vary with the number of nodes $n$. Compared to Geometric Retransmission ($K=1$), whose stable region rapidly diminishes as $n$ increases, Exponential Backoff is much more robust, and more resilient in dealing with transient fluctuations of traffic.

## APPENDIX I. DERIVATION OF $S_L^{K-Exp}$ AND $\hat{\lambda}_{\max\_S_L}^{K-Exp}$ OF *K*-EXPONENTIAL BACKOFF ($1<K<\infty$)

The lower bound $q_l^{K-Exp}$ is the root of equation $\rho=1$ when the probability of success $p$ converges to the desired stable point $p_L$. Let $x=(1-p_L)/q$. According to (6), the offered load $\rho$ can be written as

$$\rho = \lambda\left(\frac{1}{1-x} - \left(\frac{1}{1-x} - \frac{1}{p_L}\right) \cdot x^K\right) = \lambda\left(\frac{1-x^K}{1-x} + \frac{1}{p_L}x^K\right). \quad (61)$$

Suppose that $x^*$ is the solution of equation $\rho=1$. From (61), we have

$$p_L \frac{1-x^K}{1-x} + x^K = \frac{n}{\hat{\lambda}} \cdot p_L. \quad (62)$$

With a large $n$, $x^*>1$, and we have

$$p_L \frac{1-x^K}{1-x} + x^K \approx x^K. \quad (63)$$

Substituting (63) into (62), we can obtain $x^*$ approximately as follows

$$x^* \approx (np_L/\hat{\lambda})^{1/K}. \quad (64)$$

The lower bound $q_l^{K-Exp}$ is therefore given by

$$q_l^{K-Exp} = (1-p_L)/x^* \approx \frac{1-p_L}{\sqrt[K]{np_L/\hat{\lambda}}}. \quad (65)$$

The absolute-stable region $S_L^{K-Exp}$ can be then obtained by combining (24) and (65).

With a large $K$, we have

$$q_l^{K-Exp} \approx \frac{\hat{\lambda}^{1+1/K}}{\sqrt[K]{n}} \approx \frac{\hat{\lambda}}{\sqrt[K]{n}}. \quad (66)$$

$\hat{\lambda}_{\max\_S_L}^{K-Exp}$ can be obtained by combining (24) and (66):

$$\hat{\lambda}/n^{1/K} = -\ln p_S/n \Rightarrow \hat{\lambda}_{\max\_S_L}^{K-Exp} \approx \ln n^{1-1/K}/n^{1-1/K}. \quad (67)$$

## APPENDIX II. UNDESIRED STABLE POINT OF EXPONENTIAL BACKOFF $p_A^{Exp}$

Let

$$x = n(1/q-1)/p_A^{Exp}. \quad (68)$$

According to (56), we have

$$xe^x = n(1/q-1)\exp(n/q). \quad (69)$$

Since $n(1/q-1)\exp(n/q)>0$, $x$ can be uniquely represented as:

$$x = W_0(n(1-q)/q \cdot \exp(n/q)). \quad (70)$$

By applying the monotonic increasing property of $W_0(z)$ and the property of $W_0(xe^x)=x$ to the following inequality:

$$(n/q-n)\exp(n/q-n) \leq (n/q-n)\exp(n/q) \leq (n/q)\exp(n/q), \quad (71)$$

we immediately obtain

$$n/q - n \leq x \leq n/q. \quad (72)$$

Suppose that

$$x = n/q - \delta, \quad (73)$$

for some $0\leq\delta\leq n$. According to (70) and (73), we have

$$(n/q-\delta)\cdot\exp(n/q-\delta) = (n/q-n)\exp(n/q). \quad (74)$$

It follows that

$$\exp(\delta) = (n/q-\delta)/(n/q-n). \quad (75)$$

Since $n/q \gg \delta$ for large $n$, $\delta$ can be approximately given by

$$\delta \approx -\ln(1-q). \quad (76)$$

from (75).

Finally, by combining (73) and (76), we have

$$x \approx n/q + \ln(1-q). \quad (77)$$

The undesired stable point $p_A^{Exp}$ can be then obtained from (68) and (77) as

$$p_A^{Exp} = n(1/q-1)/x \approx \frac{n(1-q)}{n+q\ln(1-q)} \approx 1-q. \quad (78)$$

## APPENDIX III. CHARACTERISTICS OF UNDESIRED STABLE POINT $p_A$ WITH CUTOFF PHASE $1<K<\infty$

1) $p_A$ is monotonic increasing with respect to cutoff phase $K$;
2) For any given retransmission factor $q$, $p_A\to 0$ as $n\to\infty$.

Proof: 1) It can be seen from (45) that for any given $p_t$ and $q$, $g(p_t)$ is monotonic increasing with respect to the cutoff phase $K$. Suppose that $K_1<K_2$. Let $p_{A,1}$ and $p_{A,2}$ represent the corresponding convergent points of $p_t$ if $p_t<p_S$. According to (44) we have

$$p_{A,1} = \exp\{-n/g_{K_1}(p_{A,1})\}, \quad (79)$$

and

$$p_{A,2} = \exp\{-n/g_{K_2}(p_{A,2})\}. \quad (80)$$

Combining (79) and (80), we have

$$g_{K_2}(p_{A,2})\ln\frac{1}{p_{A,2}} = n = g_{K_1}(p_{A,1})\ln\frac{1}{p_{A,1}} < g_{K_2}(p_{A,1})\ln\frac{1}{p_{A,1}}. \quad (81)$$

Since both $g(p)$ and $\ln(1/p)$ are monotonic decreasing functions of $p$, the inequality (81) can only hold for $p_{A,1}<p_{A,2}$. It follows that $p_A$ is monotonic increasing with respect to $K$.

2) According to (46), we know that $p_A$ satisfies

$$p_A = \exp\left\{-n \Big/ \left(\frac{p_A q}{p_A + q - 1} - \left(\frac{p_A q}{p_A + q - 1} - 1\right) \cdot \left(\frac{1-p_A}{q}\right)^K\right)\right\}. \quad (82)$$

Let $x=(1-p_A)/q$, and rewrite (82) as

$$-\ln(1-qx) \cdot \left((1-qx)\frac{1-x^K}{1-x} + x^K\right) = n. \quad (83)$$

For any finite cutoff phase *K*, it is clear from (83) that $x \to 1/q$ as $n \to \infty$. Therefore, the undesired stable point $p_A \to 0$ as $n \to \infty$. □